\documentclass[12pt]{iopart}

\usepackage{iopams}
\usepackage{amssymb}
\usepackage{amsfonts}
\usepackage{mathrsfs}
\usepackage{bbm,bm}
\usepackage{graphicx}
\usepackage{xcolor}
\usepackage{hyperref}
\hypersetup{colorlinks=true,breaklinks=true,pdfstartview=FitH,linkcolor=blue,citecolor=blue}

\newcommand{\bra}[1]{\left\langle #1\right\vert}
\newcommand{\ket}[1]{\left\vert #1\right\rangle}

\begin{document}
\title{Geometric picture of quantum discord for two-qubit quantum states}

\author{Mingjun Shi$^1$, Fengjian Jiang$^1$, Chunxiao Sun$^1$ and Jiangfeng Du$^{1,2}$}

\address{$^1$ Department of Modern Physics, University of Science and
Technology of China, Hefei, Anhui 230026, People's Republic of China}
\address{$^2$ Hefei National Laboratory for Physical Sciences at Microscale \&
Department of Modern Physics, University of Science and Technology of China,
Hefei, Anhui 230026, People's Republic of China}

\ead{\mailto{shmj@ustc.edu.cn}, \mailto{djf@ustc.edu.cn}}

\begin{abstract}
Among various definitions of quantum correlations, quantum discord has attracted considerable attention.
To find analytical expression of quantum discord is an intractable task.
Exact results are known only for very special states, namely, two-qubit X-shaped states.
We present in this paper a geometric viewpoint, from which two-qubit quantum discord can be described clearly. The known results about X state discord are restated in the directly perceivable geometric language. As a consequence, the dynamics of classical correlations and quantum discord for an X state in the presence of decoherence is endowed with geometric interpretation.
More importantly, we extend the geometric method to the case of more general states, for which numerical as well as analytical results about quantum discord have not been found yet. Based on the support of numerical computations, some conjectures are proposed to help us establish geometric picture.
We find that the geometric picture for these states has intimate relationship with that for X states. Thereby in some cases analytical expressions of classical correlations and quantum discord can be obtained.
\end{abstract}

\pacs{03.65.Ta, 03.67.-a}

\maketitle

\section{Introduction}

Correlation is the relationship between different things, and is a pervasive phenomena in nature. It is the way by which we learn the external world, and is the bridge on which we communicate each other and transmit information from this end to the other.

In the classical world, correlations have been well studied from the viewpoint of information theory (see, for example, \cite{Cover}).
However ``quantizing'' classical information or correlation is definitely not an effortless work. The difficulties crop up in that quantum information, unlike the classical counterpart, is encoded in quantum states which may not be orthogonal and thus may not be distinguished unambiguously, and moreover, quantum systems can be correlated in ways inaccessible to classical objects.

One of the prominent features of quantum correlation is entanglement. Entangled states cannot be prepared with the help of local operations and classical communication (LOCC) and thus they are nonclassical. Entanglement is indeed an important aspect of quantum correlation and is a prerequisite for many tasks of quantum information processing \cite{Horodecki2009RMP}. Nevertheless, entanglement is not the only aspect of quantum correlation, and the notion of quantum correlation is more general than entanglement. For example, there exists quantum nonlocality without entanglement
\cite{Bennett1999PRA,Bennett1999PRL,Horodecki2003PRL}.

Various approaches, other than through entanglement, have been proposed to study the correlations in composite quantum system. The first attempt at quantifying quantum contents of correlations is due to Zurek. The concept of quantum discord is proposed and develops to a measure of how non-classical the underlying correlation of two quantum systems is \cite{Zurek2003RMP,Zurek2000AnnPhys,Ollivier2001PRL}.
The important issue is the existence of quantum correlations beyond entanglement in separable states.
Quantifying classical correlations in a bipartite quantum state and splitting the total correlation into a classical and a quantum part was presented by Henderson and Vedral \cite{Henderson2001JPA}. Later the Henderson-Vedral (H-V) classical correlation was shown to have an operational meaning: The regularization of H-V classical correlation is just the maximal amount of common random bits obtained by one-way LOCC operations in excess of communication invested \cite{Devetak2004IEEE,Devetak2005PRA}.
In \cite{Oppenheim2002PRL} Oppenheim \textit{et al} presented an operational proposal, which comes from thermodynamical consideration, to quantify quantum correlations (see also \cite{Horodecki2005PRA}).
By considering the amount of noise required to erase the correlation, Groisman \textit{et al} gave an operational definition of the quantum, classical, and total amounts of correlations in a bipartite quantum state \cite{Groisman2005PRA}.
Recently Modi \textit{et al} proposed a unified view of quantum and classical correlations
\cite{Modi2010PRL}.

The above considerations shed new light on the properties of the correlation incorporated in composite quantum system. Following the division of total correlation into classical and quantum part, many works have been devoted to study the roles played by different types of correlation in quantum processes, and reveal the relationship between them. These studies involve fuzzy measurement \cite{Vedral2003PRL}, mixed-state quantum computation speedups \cite{Datta2007PRA,Datta2008PRL}, broadcasting of quantum state \cite{Piani2008PRL}, complete positivity of dynamics \cite{Rodriguez2008JPA,Shabani2009PRL}, complementarity and monogamy relationship between classical and quantum correlations
\cite{Oppenheim2003PRA,Badziag2003PRL,Koashi2004PRA} and dynamics of discord
\cite{Maziero2009PRA,Mazzola2010PRL}.

However, there is no effective method to calculate the exact results of quantum discord and other measures of quantumness analytically. Unlike the measure of entanglement, the new paradigms of quantumness of correlations are measurement oriented. What should be done in these paradigms is to extract information about system A by measuring another system B.
Given a bipartite quantum system in the state $\rho^{AB}$, when measuring system B gives the outcome $k$ with probability $p_k$, system A would be in some postmeasurement state $\rho^A_k$. For a complete measurement on system B, the $\rho_k$ and the $p_k$ are the members and probabilities of an ensemble of the local state of system A, that is, $\rho^A=\sum_kp_k\rho_k^A$.
The accessible information about system A with respect to the particular measurement is given by $S(\rho^A)-\sum_kp_kS(\rho_k^A)$, where $S(\rho)$ is the von Neumann entropy of a state $\rho$.
The major obstacle is to maximize the accessible information, or equivalently, to minimize the average entropy $\overline{S}^A=\sum_kp_kS(\rho_k^A)$, over all possible complete measurements performed on system B.
The explicit analytical results of quantum discord are known only for very special cases: Bell-diagonal states \cite{Luo2008PRA}, X-shaped states \cite{Ali2010PRA} of two-qubit system, and Gaussian states of continuous variable systems \cite{Giorda2010PRL}.

Considering this problem, we propose in this paper a geometric method to describe the quantum discord of two-qubit quantum states.
The geometric method is based on the idea of quantum steering ellipsoid which is defined in \cite{Verstraete}.
Quantum steering ellipsoid is such an ellipsoid in three-dimensional real space $\mathbb{R}^3$ that each point in the interior or on the surface represents a postmeasurement state of one qubit when particular measurement has been performed on the other qubit. We denote the quantum steering ellipsoid by $\mathfrak{E}$.
The available postmeasurement states are constrained by the $\mathfrak{E}$. Or in other words, the decomposition of one local state, say $\rho^A$, can only be performed in the $\mathfrak{E}$ (including the surface), namely, $\rho^A=\sum_kp_k\rho_k^A$ for each $\rho_k^A\in\mathfrak{E}$.
We call a postmeasurement ensemble optimal if this ensemble can minimizes the average entropy. We also call the optimal ensemble as the optimal decomposition of the local state. It can be shown that the optimal ensemble can only be found on the surface of $\mathfrak{E}$.
This situation can be compared with the optimal signal ensembles studied in \cite{Schumacher2001PRA}, where the output states of a noisy quantum channel are restricted in a convex set $\mathcal{A}$, and the optimal signal ensemble $\{p_k^{\mathrm{opt}},\rho_k^{\mathrm{opt}}\}$ is such that the Holevo quality, $\chi=S(\rho)-\sum_kp_kS(\rho_k)$ with each $\rho_k\in\mathcal{A}$ and $\rho=\sum_kp_k\rho_k$, reaches the maximum on $\{p_k^{\mathrm{opt}},\rho_k^{\mathrm{opt}}\}$.

For two-qubit X states, there are only two candidates for the optimal ensemble. We call them equi-entropy decomposition and quasi-eigendecomposition respectively. The geometric picture of these two forms of decomposition is clear: equi-entropy decomposition corresponds to a horizontal line segment, while quasi-eigendecomposition to a vertical one.
Then the known results for Bell-diagonal states and X states can be ``seen'' in this picture.
Subsequently, we study the dynamics of classical correlations and quantum discord in the presence decoherence. In \cite{Mazzola2010PRL}, it has been shown that there is a sudden transition from classical to quantum decoherence regime for some Bell-diagonal states undergoing non-dissipative decoherence. We generalize this result to the case of general X states. The sudden transition can even be ``seen'' in the geometric picture.

Not only can the geometric method be used to recover the known results, but also it should help us seek the possible analytical expressions of quantum discord for more general states. Following this line of thought, we consider a class of two-qubit states that have more complicated forms than X states. For these states, we can not give a thoroughly analytical procedure to derive the classical correlations or quantum discord. However, numerical computations give us the interesting results. We find that, just like the case of X states, there are only two possibilities as to the optimal postmeasurement ensemble. One is equi-entropy decomposition, and the other, although not the quasi-eigendecomposition, has intimate relation with the quasi-eigendecomposition. In the former case, we can write out analytical expressions for classical correlations and quantum discord, while for the latter further research is needed to characterize its property. We think that these phenomena revealed by numerical work should not be accidental coincidences.
If these phenomena can be verified analytically, they will give us a geometric insight into the quantum discord.

In Section 2 we give a brief overview of the concepts of classical correlations and quantum discord. In Section 3 we introduce a very useful tool, quantum steering ellipsoid, for evaluating quantum discord of two-qubit states. We present in Section 4 the geometric picture to evaluate and describe the quantum discord of X states. For X states undergoing decoherence, the dynamics of classical correlations and quantum discord is studied and depicted geometrically in Section 5. More general states are considered in Section 6 and 7. Several conjectures and numerical tests are presented therein. Section 8 concludes.

Throughout this paper, the logarithm has base $2$. Numerical computations are performed by using Mathematica $8.0$.

\section{Classical correlations and quantum discord}

Consider a bipartite quantum system composed of particle A and particle B, which are possessed by Alice and Bob respectively. The state of the whole system is described by a density matrix $\rho^{AB}$. Total correlation between particle A and particle B is usually measured by the mutual information, that is,
\begin{equation}\label{def:total correlation}
  \mathcal{I}(\rho^{AB})=S(\rho^A)+S(\rho^B)-S(\rho^{AB}),
\end{equation}
where $\rho^{A}$ and $\rho^{B}$ are local states of A and B respectively,  $\rho^{A(B)}=\Tr_{B(A)}(\rho^{AB})$. Mutual information quantifies the strength of the correlation. For product state $\rho^{AB}=\rho^A\otimes\rho^B$, the entropy $S(\rho^{AB})$ is additive, namely, $S(\rho^{AB})=S(\rho^A)+S(\rho^B)$, and it follows that the mutual information for any product state is zero. For maximally entangled state, such as $\frac{1}{\sqrt{d}}\sum_{i=0}^{d-1}\ket{i}_A\otimes\ket{i}_B$ with $d$ the dimension of the Hilbert space of subsystem A or B, the mutual information reaches its maximal value, $2\log_2 d$. It is shown that quantum mutual information is just the minimal rate of randomness that is required to completely erase all the correlations in $\rho^{AB}$ \cite{Groisman2005PRA}.

The total correlation $\mathcal{I}$ can be split into quantum part $\mathcal{Q}$ and classical part $\mathcal{C}$, namely, $\mathcal{I}=\mathcal{Q}+\mathcal{C}$.
There are several ways to define the measure of classical correlations. Here we adopt the definition given by Henderson-Vedral \cite{Henderson2001JPA}, which quantifies the information gained about one subsystem from the measurement on the other.

Suppose that Bob performs POVM measurements on his particle B. The set of POVM elements is denoted by $\mathcal{M}=\{M_k\}$ with $M_k\geqslant 0$ and $\sum_kM_k=\mathbbm{1}$. The probability that outcome $k$ is obtained is given by
\begin{equation*}
    p_k=\Tr[\rho^{AB}(\mathbbm 1\otimes M_k)].
\end{equation*}
The postmeasurement state of particle A that corresponds to the outcome $k$ is
\begin{equation} \label{rho Ak}
   \rho^A_k=\frac{1}{p_k}\Tr_B[\rho^{AB}(\mathbbm 1\otimes M_k)].
\end{equation}
Considering all POVM elements $M_k$'s, the postmeasurement states of particle A are characterized by the ensemble $\{p_k,\rho_k^A\}$. Note that Alice's local state $\rho^A$ remains unchanged, namely, $\rho^A=\sum_k p_k\rho_k^A$ for any postmeasurement ensemble. Or in other words, Bob's POVM measurements induce a decomposition of Alice's local state $\rho^A$
into the ensemble $\{p_k,\rho_k^A\}$.

The information about particle A that is acquired by Bob's specific POVM $\mathcal{M}$ is given by
\begin{equation*}
    S(\rho^A)-\sum_k p_k S(\rho_k^A).
\end{equation*}
The dependence on the measurement procedure can be removed by maximization over all possible POVMs.
The classical correlation is then defined as
\begin{equation}\label{def:classical correlation}
    \mathcal{C}^\leftarrow=\max\limits_{\mathcal{M}}\big[S(\rho^A)-\sum_k p_k S(\rho_k^A)\big]
    =S(\rho^A)-\min_{\mathcal{M}}\sum_{k}p_k S(\rho_k^A),
\end{equation}
where the maximization and minimization are taken over all of Bob's POVM measurements. The left-arrow over $\mathcal{C}$ indicates the situation that Bob performs measurement to acquire the information about Alice's system. Similarly, if Alice performs POVM, $\mathcal{N}=\{N_j\}$, we can define the information gained about particle B by measuring particle A as
\begin{equation*}
    \mathcal{C}^\rightarrow=\max\limits_{\mathcal{N}}\big[S(\rho^B)-\sum_j p_j S(\rho_j^B)\big]
    =S(\rho^B)-\min_{\mathcal{N}}\sum_{j}p_j S(\rho_j^B).
\end{equation*}
Generally, $\mathcal{C}^{\rightarrow}\neq\mathcal{C}^{\leftarrow}$, meaning that the classical information is asymmetric.

It is natural to define quantum correlation as the difference between total correlation and classical correlation, namely,
\begin{eqnarray}
   \mathcal{Q}^{\leftarrow} & =\mathcal{I}-\mathcal{C}^{\leftarrow} \nonumber \\
   & =\min_{\mathcal{M}}\sum_{k}p_k S(\rho_k^A)+S(\rho^B)-S(\rho^{AB}). \label{def:quantum correlation}
\end{eqnarray}
Similarly for $\mathcal{Q}^{\rightarrow}=\mathcal{I}-\mathcal{C}^{\rightarrow}$.

Quantum correlation $\mathcal{Q}$ is also called quantum discord. Quantum discord, which is originally defined as the the difference between two classically identical (but quantumly distinct) formulas that measure the amount of mutual information of a pair of quantum systems \cite{Zurek2000AnnPhys,Zurek2003RMP}, aims to capture all the quantum correlations, not limited to entanglement. There is a fundamental difference between entanglement and discord for mixed states, although they are equivalent for pure states. A typical example of this is the separable states with nonvanishing discord \cite{Ollivier2001PRL}. Other forms of definition of quantum discord can be found in \cite{Brodutch2010PRA}.

To obtain quantum discord or classical correlation, one has to make considerable effort to minimize the average entropy $\overline{S}^A=\sum_kp_kS(\rho_k^A)$ over all possible measurements on particle B. In the next section, we introduce a useful tools, quantum steering ellipsoid, which will help us to establish a geometric picture about these concepts.


\section{Quantum steering ellipsoid}

First we express the states and POVM elements in Hilbert-Schmidt space.
Let $\rho^{AB}$ be a two-qubit state shared by Alice and Bob. It can be written as
$\rho^{AB}=\frac{1}{4}\sum_{\alpha,\beta=0}^{3}R_{\alpha\beta}\,\sigma_\alpha\otimes\sigma_\beta$
where $\sigma_0$ is the $2\times2$ identity matrix, $\sigma_i$ $(i=1,2,3)$ are Pauli matrices, and $R_{\alpha\beta}=\Tr[\rho^{AB}(\sigma_\alpha\otimes\sigma_\beta)]$ are all real numbers. We arrange the $16$ coefficients $R_{\alpha\beta}$ into a $4\times4$ matrix $R=(R_{\alpha\beta})$. Note that $R_{00}$ is just the trace of $\rho^{AB}$ and equal to one.
We write $M_k$, one element of Bob's POVM, as $M_k=\sum_{\alpha=0}^{3}x_{k,\alpha}\sigma_\alpha$. Similarly the state $\rho_k^A$ (see (\ref{rho Ak})) can be expressed as $\rho_k^A=\frac{1}{2}\sum_{\alpha=0}^{3}y_{k,\alpha}\sigma_\alpha$.

Let's define two four-components vectors in the row form, $\mathbf{y}_k=(y_{k,0},\,\vec{y}_k)$ and $\mathbf{x}_k=(x_{k,0},\,\vec{x}_k)$, where $\vec{y}_k=(y_{k,1},\,y_{k,2},\,y_{k,3})$ and $\vec{x}_k=(x_{k,1},\,x_{k,2},\,x_{k,3})$. Note that $y_{k,0}=1$ for all $k$ and $\vec{y}_k$ is the Bloch vector of $\rho_k^A$. Direct calculation shows the following equation.
\begin{equation}\label{relation between y and x}
  p_k\,\mathbf{y}_k=\mathbf{x}_{k}\,R^T, \quad
  p_k=\sum_{\alpha=0}^{3}R_{0\alpha}\,x_{k,\alpha},
\end{equation}
where the superscript $T$ means matrix transpose. (\ref{relation between y and x}) provides the
relationship between Bob's measurement and the corresponding components in Alice's ensemble. For
entangled states, the matrix $R$ is of full rank, and the vector $\mathbf{y}_{k}$ is in a one-to-one correspondence to the vector $\mathbf{x}_{k}$. In the following, we will treat $R$ as a full rank matrix. Some states with singular $R$ will be discussed in Section 6.

Also note that (\ref{relation between y and x}) imposes constraint on the vector $\mathbf{y}_k$: although $\mathbf{x}_k$ can represent any projective measurement, the vector $\mathbf{y}_k$ can not be arbitrary. For example, $\mathbf{y}_k$ can not represent a pure state unless $\rho^{AB}$ is a pure state. It is pointed out in \cite{Verstraete} that the allowed $\mathbf{y}_{k}$ must satisfy
\begin{equation}\label{ellipsoid}
  \mathbf{y}_k\,\big[R^{-T}\,\eta\,R^{-1}\big]\,(\mathbf{y}_k)^T\geqslant 0,
\end{equation}
where $R^{-T}=(R^{-1})^T$ and $\eta=\mathrm{diag}(1,-1,-1,-1)$. In fact, (\ref{ellipsoid}) comes from the
requirement that each $M_k$ is nonnegative.  Noting that the 3-component
vector $\vec{y}_k=(y_{k,1},\,y_{k,2},\,y_{k,3})$ is the Bloch vector of $\rho_k^A$, we can see that (\ref{ellipsoid}) describes a ellipsoidal region in three-dimensional real space. It means that, for each $\rho_k^A$ allowed to appear in Alice's ensemble $\{p_k,\rho_k^A\}$, the
corresponding Bloch vector $\vec{y}_k$ is constrained within an ellipsoid (including the surface).
The ellipsoid given by (\ref{ellipsoid}) is called ``steering ellipsoid'' in \cite{Verstraete}. We denote it by $\mathfrak{E}$.

The steering ellipsoid renders concrete geometric picture when we do the minimization of the
average entropy $\overline{S}^A$. Each point belonging to $\mathfrak{E}$ corresponds to some $\rho_k^A$. By noting that the $\mathfrak{E}$ is convex and entropy function is concave, we see that the minimal value of $\overline{S}^A$ must be attained on the surface of $\mathfrak{E}$.
For any point on the
surface of $\mathfrak{E}$, that is, for any vector $\mathbf{y}$ such that the equality in (\ref{ellipsoid}) holds, the corresponding vector $\mathbf{x}$ (see (\ref{relation between y and x})), must satisfy $x_0^2=x_1^2+x_2^2+x_3^2$. Such a vector $\mathbf{x}$ represents rank-one element of Bob's POVM that can be taken to be proportional to the one-dimensional projector $\Pi$, namely, $M=m\Pi$. With the factor $m$ absorbed into the probability $p$, we say that any point on the surface of $\mathfrak{E}$ is induced by Bob's projective measurement. Therefore in order to obtain optimal ensemble of Alice's state, Bob need only to perform projective measurements. This fact has been pointed out by Hamieh \textit{et al} in \cite{Hamieh2004PRA}. We give a geometric description here.

\section{Quantum discord of two-qubit X states}

For general two-qubit state, $\mathfrak{E}$ is too complicated to be dealt with. However for a specific class of states, called X states, we will show that the
geometric picture is very clear. In various situations X states have been used to demonstrate significant quantum phenomena, for examples, entanglement sudden death or birth \cite{Yu2006PRL,Lopez2008PRL}, dynamics of quantum and classical correlations \cite{Maziero2009PRA,Maziero2010PRA}, and sudden transition between classical and quantum decoherence \cite{Mazzola2010PRL}. The density matrix of a general two-qubit X state is
\begin{equation}\label{X state}
    \rho=\left(
    \begin{array}{cccc}
      a & 0 & 0 & u\,e^{i\mu} \\
      0 & b & v\,e^{i\nu} & 0 \\
      0 & v\,e^{-i\nu} & c & 0 \\
      u\,e^{-i\mu} & 0 & 0 & d
    \end{array}\right).
\end{equation}
where $a+b+c+d=1$ and $u,v\geqslant 0$. It is required that $u^2\leqslant ad$ and $v^2\leqslant bc$ to assure the positivity of the density matrix. By local unitary operations, the off-diagonal entries can be transformed to real ones. Since all correlations are invariant under
local unitary operations, it suffices to consider X states with all the entries of density matrix being real. However we will remain at the form given by (\ref{X state}) for later references.

The matrix $R$ is given by
\begin{equation} \label{matrix R}
  R=\left(
  \begin{array}{cccc}
    1 & 0 & 0 & a-b \\[-5pt]
      &   &   & \quad +c-d \\[5pt]
    0 & 2u\cos\mu & -2u\sin\mu & 0 \\[-5pt]
      & \quad +2v\cos\nu & \quad +2v\sin\nu &  \\[5pt]
    0 & -2u\sin\mu & -2u\cos\mu & 0 \\[-5pt]
      & \quad -2v\sin\nu & \quad +2v\cos\nu &   \\[5pt]
    a+b & 0 & 0 & a-b       \\[-5pt]
    \quad -c-d &  &   & \quad -c+d
  \end{array}
  \right)
\end{equation}
When $\det(R)=16(bc-ad)(u^2-v^2)\neq 0$, the inverse $R^{-1}$ exists. From (\ref{ellipsoid}), we can write the equation of the ellipsoid $\mathfrak{E}$, that is,
\begin{equation}\label{ellipsoid of 2-qubit state}
  \frac{{y'_1}^2}{l_1^2}+\frac{{y'_2}^2}{l_2^2}
   +\frac{{y'_3}^2}{l_3^2}=1,
\end{equation}
where $y'_1=y_1\cos\phi-y_2\sin\phi$, $y'_2=y_1\sin\phi+y_2\cos\phi$, $y_3'=y_3-Y_3$ with
$\phi=(\mu+\nu)/2$, and three major axes are given by
\begin{eqnarray}
   & l_1=\frac{u+v}{\sqrt{(a+c)(b+d)}}, \;\;
   l_2=\frac{|u-v|}{\sqrt{(a+c)(b+d)}}, \label{parameter l1 and l2}\\
   & l_3=\frac{|ad-bc|}{(a+c)(b+d)}, \;\;
   Y_3=\frac{ab-cd}{(a+c)(b+d)}. \label{parameter l3 and Y3}
\end{eqnarray}

\begin{figure}
\begin{center}
  \includegraphics[width=0.6\textwidth]{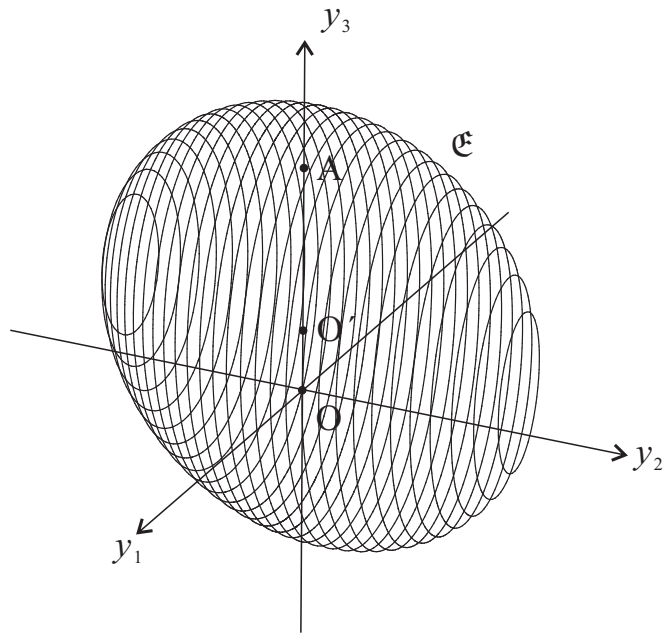}
\end{center}
  \caption{Schematic plot of quantum steering ellipsoid $\mathfrak{E}$ of a two-qubit X state. In $(O\,y_1\,y_2\,y_3)$ frame, the center of $\mathfrak{E}$, denoted by $O'$, is on the $y_3$-axis. Point $A$ represents the state of qubit A.}\label{fig:ellipsoid}
\end{figure}

In the coordinate frame $(O'\,y'_1\,y'_2\,y'_3)$, the ellipsoid $\mathfrak{E}$ takes the standard form, which comes from the form in $(O\,y_1\,y_2\,y_3)$ frame by translating along $y_3$ axis and rotating about $y_3$ axis.
In the $(O\,y_1\,y_2\,y_3)$ frame, the center of $\mathfrak{E}$ is at the point $O'$ with coordinates $(0,0,Y_3)$ (see figure \ref{fig:ellipsoid}). The Bloch vector of $\rho^A$ is given by
\begin{equation*}
    \vec{r}^A=(0,\,0,\,a+b-c-d),
\end{equation*}
which is represented by a point $A$ on the $y_3$ axis. Point $A$ can not be outside $\mathfrak{E}$ due to the fact that $a+b-c-d\in[Y_3-l_3,Y_3+l_3]$.

We now proceed to find Alice's optimal ensemble $\{p_k,\rho_k^A\}^{\mathrm{opt}}$. When Bob performs complete projective measurements, Alice's ensemble has two components, i.e., $k=1,2$, and both $\rho_1^A$ and $\rho_2^A$ are on the surface of $\mathfrak{E}$. In the following, we demonstrate a geometric picture to describe how to obtain Alice's optimal ensemble and thereby the value of $\overline{S}^A_{\min}$.

\begin{figure}[tbph]
\begin{center}
\includegraphics[width=0.5\columnwidth]{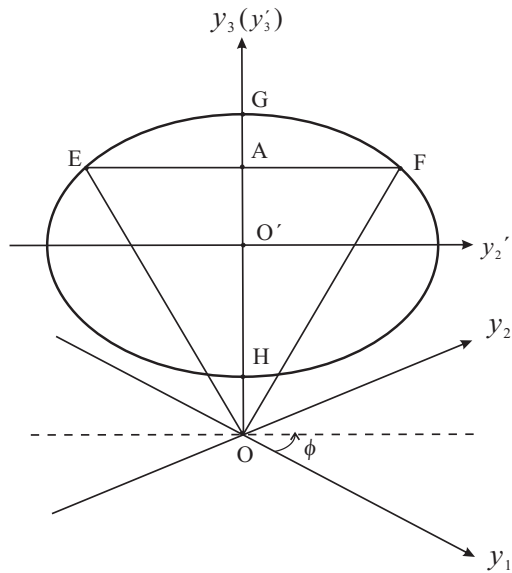}
\end{center}
\caption{Schematic plot of the largest ellipse given by $(y'_1/l_1)^2+(y'_3/l_3)^2=1$. Two pairs of points, $(E,F)$ and $(G,H)$, are the only candidates for Alice's optimal ensemble which will give the minimal value of average entropy $\overline{S}^A$.} \label{fig:LE for X states}
\end{figure}

Imagine a class of planes that contain $y'_3$ (or $y_3$) axis and intersect the ellipsoid $\mathfrak{E}$. Each section is an ellipse. By noting $l_1\geqslant l_2$, the largest section must be in the $y'_1\,y'_3$ plane. Then the largest ellipse (LE) takes the form
\begin{equation*}
    \frac{{y'_1}^2}{l_1^2}+\frac{{y'_3}^2}{l_3^2}=1.
\end{equation*}
The $y_3$ axis
intersects LE at two points $G$ and $H$. Consider a line parallel to $y'_1$ axis and passing through point
$A$. This line will intersect the LE at two points $E$ and $F$ (see figure \ref{fig:LE for X states}). Remember that point $A$ stands
for Alice's local state $\rho^A$. Then $\rho^A$ can be expressed as either of the two forms of convex sum:
$\rho^A=p_G\rho_G+p_H\rho_H$ or $\rho^A=p_E\rho_E+p_F\rho_F$, where $\rho_G$ is the state corresponding to the point $G$ and the probability $p_G=AH/GH$, and similarly for other states and probabilities. These two forms of convex sum lead us to the average entropies,
\begin{eqnarray}
    & S_{GH}=p_GS(\rho_G)+p_HS(\rho_H), \label{average entropy GH} \\
    & S_{EF}=p_ES(\rho_E)+p_FS(\rho_F)=S(\rho_E)=S(\rho_F). \label{average entropy EF}
\end{eqnarray}
We call (\ref{average entropy GH}) quasi-eigendecomposition, meaning that $\rho^A$, $\rho_G$ and $\rho_H$ have the same eigenstate. We call (\ref{average entropy EF}) equi-entropy decomposition, meaning that $\rho^A=p_E\rho_E+p_F\rho_F$ with $S(\rho_E)=S(\rho_F)$.

Now we state our main result. With Bob performing POVM measurement on his particle, the optimal postmeasurement ensemble of Alice's state is given by $\{p_G,\rho_G;\,p_H,\rho_H\}$ or $\{p_E=1/2,\rho_E;\,p_F=1/2,\rho_F\}$, and the minimal value of $\overline{S}^A$ is
\begin{equation}\label{eq:min SA}
  \overline{S}_{\min}^A=\min\{S_{GH},\;S_{EF}\}.
\end{equation}
It follows that the classical correlation and quantum discord are given respectively by
\begin{eqnarray*}
  & \mathcal{C}^{\leftarrow}(\rho)=S(\rho^A)-\min\{S_{GH},\,S_{EF}\}, \\
  & \mathcal{Q}^{\leftarrow}(\rho)=\min\{S_{GH},\,S_{EF}\}+S(\rho^B)-S(\rho^{AB}).
\end{eqnarray*}

The result (\ref{eq:min SA}) can be derived by using the conclusion in \cite{Ali2010PRA}. As pointed
out in \cite{Ali2010PRA}, there are two candidates for Bob's measurements which will induce Alice's optimal ensemble. In our notations, these candidates are denoted by 4-component vectors $\mathbf{x}_{\pm}$: (i)
$\mathbf{x}_{\pm}=\big(\frac{1}{2},0,0,\pm\frac{1}{2}\big)$; (ii)
$\mathbf{x}_{\pm}=\big(\frac{1}{2},\pm\frac{1}{2}\sin\theta,\pm\frac{1}{2}\cos\theta,0\big)$, where the measurement parameter $\theta$ will be determined latter.

For case (i), it follows from (\ref{relation between y and x}) that
\begin{eqnarray*}
    & p_+=a+c, \quad \mathbf{y}_{+}=\bigg(1,\;0,\;0,\;\frac{a-c}{a+c}\bigg), \\[10pt]
    & p_-=b+d, \quad \mathbf{y}_{-}=\bigg(1,\;0,\;0,\;\frac{b-d}{b+d}\bigg).
\end{eqnarray*}
It is easy to see that this case results in the two points $G$ and $H$ in figure \ref{fig:LE for X states}.

For case (ii), we have $p_+=p_-=1/2$ and $\mathbf{y}_{\pm}=(1,\vec{y}_{\pm})$
with Bloch vectors $\vec{y}_{\pm}$ given by
\begin{equation*}
  (\vec{y}_{\pm})^T=\left(
  \begin{array}{c}
    \pm2u\sin(\theta-\mu)\pm2v\sin(\theta+\nu) \\[5pt]
    \mp2u\cos(\theta-\mu)\pm2v\cos(\theta+\nu) \\[5pt]
    a+b-c-d
  \end{array}\right).
\end{equation*}
Because $|\vec{y}_{+}|=|\vec{y}_{-}|$, the entropy of the corresponding state is equal to each
other, namely, $S(\rho_+)=S(\rho_-)$. It follows that the average entropy is given by
$p_+S(\rho_+)+p_-S(\rho_-)=S(\rho_+)=S(\rho_-)$. To obtain classical correlation, we will maximize
$|\vec{y}_{+}|$ over the parameter $\theta$. In fact the maximal value of $|\vec{y}_{+}|$ is
attained when $\theta=(\pi+\mu-\nu)/2$. In this situation, Bloch vectors $\vec{y}_{\pm}$ is given by
\begin{equation*}
    \vec{y}_{\pm}=\Big(\pm2(u+v)\cos\phi,\,\mp2(u+v)\sin\phi,\,a+b-c-d\Big),
\end{equation*}
where $\phi=(\mu+\nu)/2$. It is
straightforward to check that $\vec{y}_{+}$ and $\vec{y}_{-}$ just correspond to the vector
$\overrightarrow{OF}$ and $\overrightarrow{OE}$ in figure \ref{fig:LE for X states} respectively. Thus
we have proved that the two pairs of points, $(G,H)$ and $(E,F)$, stand for the only two candidates for Alice's optimal ensemble. Now the problem of finding quantum discord for two-qubit X states is reduced to a simple
geometrical one. The only thing we have to take into account is the steering ellipsoid
$\mathfrak{E}$ and the largest ellipsoidal section.

\subsection{Bell-diagonal states}
To appreciate the geometric picture, let's consider a specific class of X states, i.e., Bell-diagonal states. Although the quantum discord of Bell-diagonal states has been calculated explicitly in \cite{Luo2008PRA}, we would like to provide a more concrete interpretation.

For a Bell-diagonal state $\rho_{\mathrm{BD}}$ given by
\begin{equation*}
    \rho_{\mathrm{BD}}=\frac{1}{4}\sum_{\mu=0}^{4}
                        t_{\mu}\,\sigma_{\mu}\otimes\sigma_{\mu},
\end{equation*}
where $t_0=1$, $1\pm t_3\geqslant|t_1\mp t_2|$ and it is assumed that $t_1t_2t_3\neq 0$,
the steering ellipsoid $\mathfrak{E}$ has the standard form, that is
\begin{equation*}
    \frac{y_1^2}{t_1^2}+\frac{y_2^2}{t_2^2}+\frac{y_3^2}{t_3^2}=1.
\end{equation*}
Moreover Alice's local state, represented by point A, coincides with the origin point $O$. Without loss of generality, we assume $|t_1|\geqslant|t_2|\geqslant|t_3|$. It is not difficult to see that the LE is given by
\begin{equation*}
    \frac{y_1^2}{t_1^2}+\frac{y_2^2}{t_2^2}=1.
\end{equation*}
According to the previous analysis, if Bob performs two-element POVM measurement, the minimal value of $\overline{S}^A$ is attained at the pair of points $(E,F)$ or the pair $(G,H)$. Note that $OE=OF=|t_1|$, $OG=OH=|t_2|$ and $|t_1|\geqslant|t_2|$. It follows that $\overline{S}^A_{\min}=S(\rho_E)$. Or more generally,
\begin{equation*}
    \overline{S}^A_{\min}=\min\{h(|t_1|),h(|t_2|),h(|t_3|)\},
\end{equation*}
with the function $h(x)$ defined by
\begin{equation} \label{function h}
    h(x)=-\frac{1+x}{2}\log\frac{1+x}{2}-\frac{1-x}{2}\log\frac{1-x}{2},
\end{equation}
for $x\in[0,1]$.

\section{Geometric picture of dynamics of quantum discord}

As an application of our result, let's consider the dynamics of quantum discord or classical correlation.
Recently, this problem has received considerable attention \cite{Maziero2009PRA,Mazzola2010PRL,Maziero2010PRA}. It has been shown that for some Bell-diagonal states passing through phase damping channel, the classical correlation can be unaffected by
decoherence. And more interestingly, the dynamics exhibits a sudden transition from classical to quantum decoherence regime \cite{Mazzola2010PRL}. We will show that in the geometric picture these phenomena can be ``seen'' clearly even for general X state.

When each particle of a two-qubit quantum system undergoes the phase damping process, the evolution of the state is expressed as
\begin{equation*}
    \rho_{AD}=\sum_{i,j=1}^{2}(K_i\otimes K_j)\rho(K_i\otimes K_j)^\dag,
\end{equation*}
where $K_1=\mathrm{diag}(\gamma,1)$ and $K_2=\mathrm{diag}(\sqrt{1-\gamma^2},0)$ are Kraus operators representing phase damping channel and $\gamma=e^{-\Gamma t}$ with $\Gamma$ the phase damping rate. Here we assume that qubit A and B endure the same noisy environment. At initial time $t=0$ the steering ellipsoid $\mathfrak{E}(0)$ is given by (\ref{ellipsoid of 2-qubit state}). At time $t>0$, the ellipsoid is transformed to $\mathfrak{E}(t)$, which is expressed by
\begin{equation*}
    \frac{{y'_1}^2}{(\gamma^2l_1)^2}+\frac{{y'_2}^2}{(\gamma^2l_2)^2}
    +\frac{{y'_3}^2}{l_3^2}=1.
\end{equation*}
That is, with $\gamma$ decreasing from $1$ to $0$, the radius of the ellipsoid along $y'_1$ axis and that along $y'_2$ axis decrease continuously from $l_1$ and $l_2$ respectively to zero whereas the radius along $y'_3$ axis (i.e., $y_3$ axis) remains the same. Since $l_1\geqslant l_2$, the LE is given by
$\big[y'_1/(\gamma^2 l_1)\big]^2+\big[y'_3/l_3\big]^2=1$
in the time evolution. The LE will shrink to $y'_3$ axis, namely, points $G$ and $H$ remain fixed and points $E$ and $F$ approach gradually to $y'_3$ axis. Also note that Alice's local state $\rho^A$ does not affected by phase damping and thus the point $A$ is fixed.

\begin{figure}[tbph]
\begin{center}
\includegraphics[width=0.5\columnwidth]{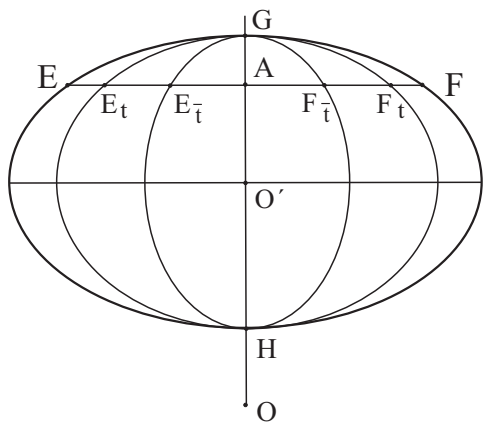}
\end{center}
\caption{Visual interpretation of dynamics of classical correlation and quantum discord.} \label{fig:dynamics LE}
\end{figure}

We now show that the dynamics of quantum discord $\mathcal{Q}^{\leftarrow}$ and classical correlation $\mathcal{C}^{\leftarrow}$ can be demonstrated clearly in the geometric picture. To this end, it suffices to consider the minimal average entropy $\overline{S}^A_{\min}$. See figure \ref{fig:dynamics LE}. There are only two possibilities with respect to the initial value of $\overline{S}^A_{\min}$.
One is that at $t=0$
Alice's optimal ensemble is determined by the points $G$ and $H$, and then $\overline{S}^A_{\min}(t=0)=S_{GH}$. At time $t$, the points $E$ and $F$ move to $E_t$ and $F_t$ respectively, while points $G$ and $H$ remain unchanged. It follows from $OE_t<OE$ that
$S_{E_tF_t}>S_{EF}>S_{GH}$. So in this case $\overline{S}_{\min}^A$ is always given by $S_{GH}$ and remains invariant. As a consequence, classical correlation $\mathcal{C}^{\leftarrow}$ does not change during the time evolution.

The other case is that initially Alice's optimal ensemble is described by two points $E$
and $F$, that is, $\overline{S}^A_{\min}(t=0)=S_{EF}<S_{GH}$. Time evolution will make $S_{E_tF_t}$ larger continuously, until the evolution reaches the
critical time, denoted by $\bar{t}$, such that $S_{E_{\bar{t}}F_{\bar{t}}}=S_{GH}$. For $t>\bar{t}$, we have $S_{E_tF_t}>S_{GH}$ and then $\overline{S}^A_{\min}(t>\bar{t})=S_{GH}$. This means that after the critical time $\bar{t}$ the classical correlation $\mathcal{C}^{\leftarrow}$ does not change any longer. In a word,
the phenomena presented in \cite{Maziero2009PRA,Mazzola2010PRL} also arise in general X states.

It is noted that the discussion presented above is not limited to phase damping channel. In fact, the geometric picture applies to any quantum channel that preserve the X form of the state, such as all unital channels in the canonical form (e.g., Pauli channel), amplitude damping channel, etc. Classical correlations may not remain constant in these more general cases.

\section{States with singular $R$}

The geometric method presented in Section 3 and 4 is based on the 3-dimensional quantum steering ellipsoid, which requires a nonsingular coefficient matrix $R$. To extend this idea to the case of singular $R$, we consider in this section two classes of states. One is the class of X states with $\det(R)=0$, which is in fact the supplement to the content of Section 4. The other is such a class of states coming from mixing two pure product states. For these states, we put forward a conjecture about the geometric description of the quantum discord.

\subsection{X states with $\det(R)=0$}

It follows from the coefficient matrix $R$ given by (\ref{matrix R}) that when $ad-bc=0$ or $u=v$, the determinant of $R$ vanishes. Recalling the ellipsoid given (\ref{ellipsoid of 2-qubit state}) and the parameters given by (\ref{parameter l1 and l2}) and (\ref{parameter l3 and Y3}), we have the following cases.

When $ad-bc=0$ and $u\neq v$, we see that $l_3=0$ and $r^A_3$ (the $y_3$-component of $\vec{r}^A$) is equal to $Y_3$. Then the ellipsoid degenerates to the ellipse. In the $(O'\,y'_1\,y'_2\,y'_3)$ frame, the equation of the ellipse is
\begin{equation*}
  \frac{{y'_1}^2}{l_1^2}+\frac{{y'_2}^2}{l_2^2}=1.
\end{equation*}
And the position of Alice's local state happens to on the origin point $O'$. Since $l_1>l_2$, the minimal value of $\overline{S}^A$ is given by
\begin{eqnarray*}
    \overline{S}^A_{\min} & =h(l_1) \\
                          & =-\frac{1+l_1}{2}\log\frac{1+l_1}{2}
                             -\frac{1-l_1}{2}\log\frac{1-l_1}{2}.
\end{eqnarray*}

When $ad-bc=0$ and $u=v\neq0$, it follows that $l_2=l_3=0$ and $r_3^A=Y_3$. We have two points given by $y'_1=l_1$ and $y'_1=-l_1$, which will determine the optimal ensemble of Alice's state.

When $ad-bc\neq0$ and $u=v\neq0$, we have $l_2=0$, and the ellipsoid degenerates to a ellipse in $y'_1\,y'_3$-plane, that is,
\begin{equation*}
    \frac{{y'_1}^2}{l_1^2}+\frac{{y'_3}^2}{l_3^2}=1.
\end{equation*}
In this case, $\overline{S}^A_{\min}$ can be easily obtained with reference to figure \ref{fig:LE for X states}.

The last case is that $u=v=0$. In this case, the density matrix $\rho$ takes the diagonal form. If $ad-bc\neq0$, the state is classically correlated and the ellipsoid reduces to the two points $G$ and $H$ in figure \ref{fig:LE for X states}, which determine the quantity $\overline{S}^A_{\min}$. If $ad-bc=0$, the state is a trivial product state.

\subsection{Mixture of two pure product states}

In this subsection, we will find the quantum and classical correlations in such states that can be written as
\begin{equation*}
  \rho=\lambda\ket{\psi_1}\bra{\psi_1}\otimes\ket{\psi_2}\bra{\psi_2}
       +(1-\lambda)\ket{\phi_1}\bra{\phi_1}\otimes\ket{\phi_2}\bra{\phi_2},
\end{equation*}
where $\lambda\in[0,1]$, and $\ket{\psi_i}$ and $\ket{\phi_i}$ ($i=1,2$) are the states of particle A and B respectively. Since the correlations remain invariant under local unitary transformations, it suffices to consider the states with the following form.
\begin{equation}\label{mixture of two PP}
  \rho=\lambda\ket{0}\bra{0}\otimes\ket{0}\bra{0}
       +(1-\lambda)\ket{\psi}\bra{\psi}\otimes\ket{\varphi}\bra{\varphi},
\end{equation}
where $\ket{\psi}=\cos\alpha\ket{0}+\sin\alpha\ket{1}$,
and $\ket{\varphi}=\cos\beta\ket{0}+\sin\beta\ket{1}$ with $\alpha,\beta\in[0,\frac{\pi}{2}]$.

A special case of (\ref{mixture of two PP}), where both $\ket{\psi}$ and $\ket{\varphi}$ are set to be $\ket{+}=\frac{1}{\sqrt2}(\ket{0}+\ket{1})$, is discussed in \cite{Henderson2001JPA} (see also \cite{Hamieh2004PRA}), and numerical evaluation is performed to inquire about the classical correlation therein. Here we consider a more general case, and will put forward a conjecture about the exact value of quantum discord or classical correlation from a geometric viewpoint.

Suppose that Bob performs POVM measurement on his qubit to acquire information about Alice's qubit. As stated earlier, we need only consider projective measurements. Let Bob's measurement operators be $M_{+}$ and $M_{-}$, that is,
\begin{equation}
  M_\pm=\frac{\mathbbm{1}}{2}\pm x_1\sigma_x\pm x_2\sigma_y\pm x_3\sigma_z,
\end{equation}
with $x_1^2+x_2^2+x_3^2=1/4$. Bob's measurement will give the result ``$+$'' with probability $p_+$ and the result ``$-$'' with probability $p_-$. The corresponding postmeasurement states of qubit A are $\rho^A_+$ and $\rho^A_-$ respectively.

\begin{figure}[tbph]
\begin{center}
\includegraphics[width=0.8\columnwidth]{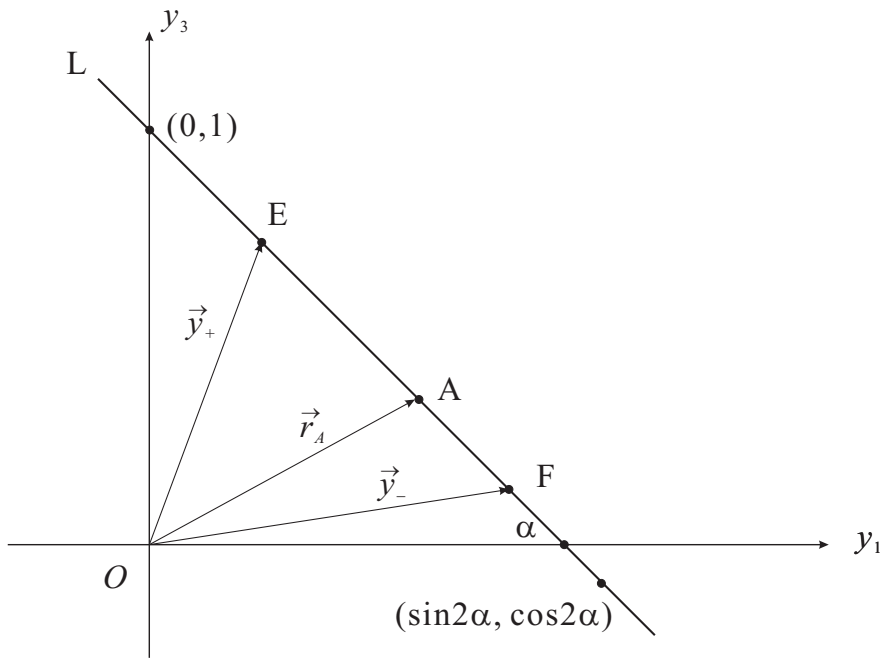}
\end{center}
 \caption{Geometric picture about the states given by (\ref{mixture of two PP}). Point $E$ and $F$ are on the line $L$ and denote the postmeasurement states of qubit A. Point $A$ denotes the local state of qubit A. Line segment $EF$ can slide between point $(0,1)$ and point $(\sin2\alpha,\cos2\alpha)$.}\label{fig:line}
\end{figure}

Define two 4-component vector $\mathbf{x}_+$ and $\mathbf{x}_-$ as
\begin{equation*}
    \mathbf{x}_{\pm}=\Big(\frac{1}{2},\,\vec{x}\Big)
      =\Big(\frac{1}{2},\,\pm x_1,\,\pm x_2,\,\pm x_3\Big).
\end{equation*}
Denote by $\vec{y}_{\pm}=(y_{\pm,1},\;y_{\pm,2},\;y_{\pm,3})$ the Bloch vectors of $\rho^A_\pm$ respectively. Then from (\ref{relation between y and x}) we have,
\begin{eqnarray}
  & p_{\pm}=\frac{1}{2}\pm x_1(1-\lambda)\sin 2\beta
            \pm x_3\big[\lambda+(1-\lambda)\cos2\beta\big], \\[10pt]
  & y_{\pm,1}=\frac{1}{p_{\pm}}\bigg\{\frac{1}{2}(1-\lambda)\sin 2\alpha
               \nonumber\\
  &      \qquad\qquad \pm x_1(1-\lambda)\sin 2\alpha\sin 2\beta
             \pm x_3(1-\lambda)\sin 2\alpha\cos 2\beta\bigg\}, \label{y1 of E and F}\\[10pt]
  & y_{\pm,2}=0, \label{y2 of E and F}\\[10pt]
  & y_{\pm,3}=\frac{1}{p_\pm}\Bigg\{\frac{1}{2}[\lambda+(1-\lambda)\cos 2\alpha]
               \nonumber\\
  & \qquad\qquad  \pm x_1(1-\lambda)\cos 2\alpha\sin 2\beta
                  \pm x_3[\lambda+(1-\lambda)\cos 2\alpha\cos 2\beta]\bigg\}
                  \label{y3 of E and F}.
\end{eqnarray}
We can see from these expressions that
\begin{equation}\label{line}
  y_{\pm,3}+y_{\pm,1}\tan\alpha=1.
\end{equation}
(\ref{line}) means that the two points corresponding to Bloch vector $\vec{y}_+$ and $\vec{y}_-$ are located on the line $L$ that lies in $y_1\,y_3$ plane and passes through the point $(0,1)$ with the slope $-\tan\alpha$. Note that the Bloch vector of $\rho^A$ is given by
\begin{equation*}
    \vec{r}^A=\Big((1-\lambda)\sin2\alpha,\;0,\;\lambda+(1-\lambda)\cos2\alpha\Big).
\end{equation*}
Then the point $A$, denoting the state $\rho^A$, is also on the line $L$. See figure \ref{fig:line}.

Before using this picture to find the quantum discord of the state given by (\ref{mixture of two PP}), let's add some remarks.

Roughly speaking, the set of postmeasurement states of qubit A (i.e., $\rho^A_+$ and $\rho^A_-$) are restricted on the line $L$. For convenience, we denote the two states by point $E$ and $F$ respectively in figure \ref{fig:line}. Assume that $E$ is on the left side of point $A$ and $F$ on the right side of $A$. Then it should be noted that $E$ can not be located on the left side of point $(0,1)$, because the length of $OE$ can not be larger than one. For the same reason, $F$ can not be on the right side of point $(\sin2\alpha,\cos2\alpha)$. Then the line segment $EF$, which represents the set of all available postmeasurement states of qubit A, slides along line $L$ between point $(0,1)$ and point $(\sin2\alpha,\cos2\alpha)$. This picture is somewhat different from that presented in Section 4, where the steering ellipsoid takes up a fixed region for given state and does not depend on the choice of the measurements performed by Bob on qubit B. But here the line segment $EF$ is ``moving'', in the sense that both the length and the position of $EF$ depend on Bob's measurements.

Now we propose a conjecture about minimal average entropy $\overline{S}^A_{\min}$.
Given a two-qubit state (\ref{mixture of two PP}), Bob performs two-element POVM measurement on qubit B. If Alice's postmeasurement ensemble $\{p_k,\rho_k^A\}_{k=+,-}$ minimizes the average entropy $\overline{S}^A$, then $S(\rho^A_+)=S(\rho^A_-)$.

\begin{figure}[tbph]
\begin{center}
\includegraphics[width=0.9\columnwidth]{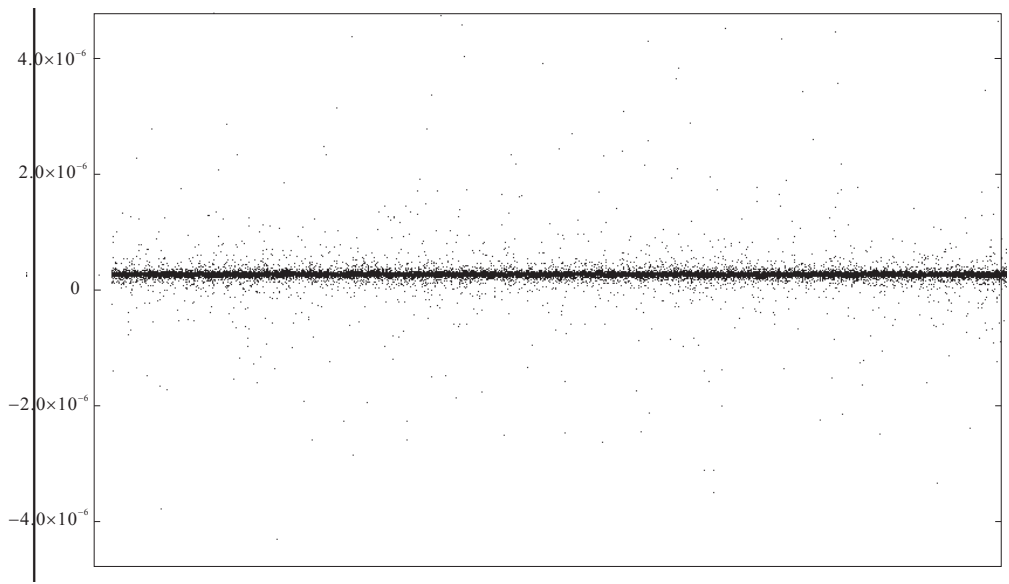}
\end{center}
\caption{Numerical test of the conjecture that $OE=OF$ is the necessary condition for $\{\rho_E,\rho_F\}$ to be the optimal ensemble of $\rho^A$. $1.5\times10^5$ random states are tested and the values of $|OE-OF|$ are calculated. The results show that almost  $|OE-OF|\ll 10^{-6}$.}\label{fig:numerical proof}
\end{figure}

In other words, $OE=OF$ is the necessary condition which must be satisfied in order that the average entropy $\overline{S}^A$ takes the minimal value. To test this conjecture, we select randomly $1.5\times10^5$ two-qubit states with form given by (\ref{mixture of two PP}), and for each state calculate numerically the value of $\overline{S}^A_{\min}$ and the corresponding measurement parameters, namely, $x_i$ with $i=1,2,3$. From (\ref{y1 of E and F}), (\ref{y2 of E and F}) and (\ref{y3 of E and F}), we get the coordinates of point $E$ and $F$ and also the length of $OE$ and $OF$. In figure \ref{fig:numerical proof}, we plot the value of $OE-OF$ for the $1.5\times10^5$ states. We see that $|OE-OF|\ll0$. Numerical results confirm our conjecture.

If the conjecture is indeed true, we need only consider the situation that $S(\rho^A_+)=S(\rho^A_-)$ or $OE=OF$. It follows that
\begin{equation*}
    \overline{S}^A=p_+S(\rho^A_+)+p_-S(\rho^A_-)=S(\rho^A_+)=S(\rho^A_-).
\end{equation*}
To obtain $\overline{S}^A_{\min}$, we need only to maximize $OE$ or $OF$ under the condition that $OE=OF$. It is not a difficult work. Following this line of thought, we obtain the classical correlation and quantum discord of states (\ref{mixture of two PP}) and plot the results in figures.

\begin{figure}
\begin{center}
  \includegraphics[width=\textwidth]{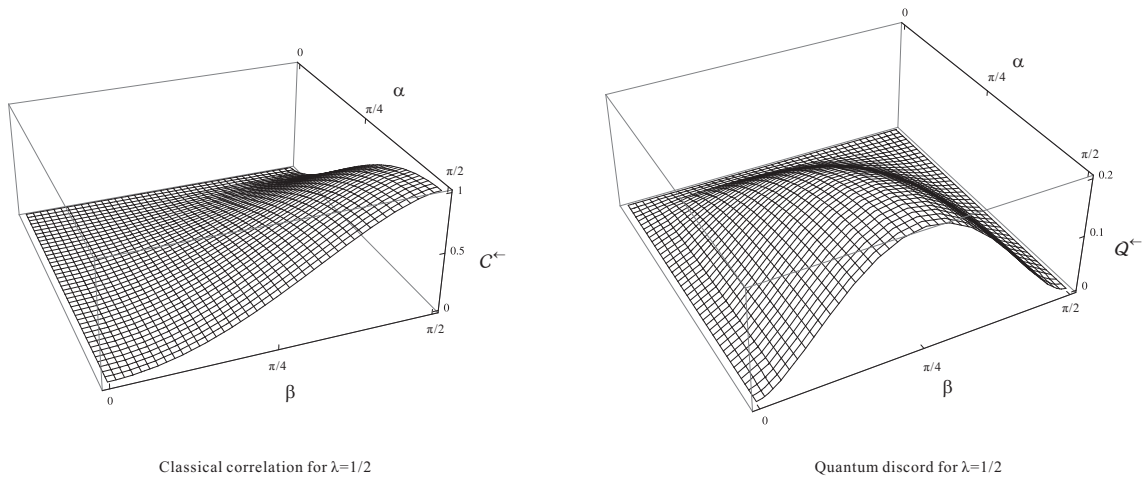}
\end{center}
 \caption{Classical correlation $\mathcal{C}^{\leftarrow}$ and quantum discord $\mathcal{Q}^{\leftarrow}$ of the states given by (\ref{mixture of two PP}) with $\lambda=\frac{1}{2}$.} \label{fig:CandQ1}
\end{figure}

\begin{figure}
\begin{center}
  \includegraphics[width=\textwidth]{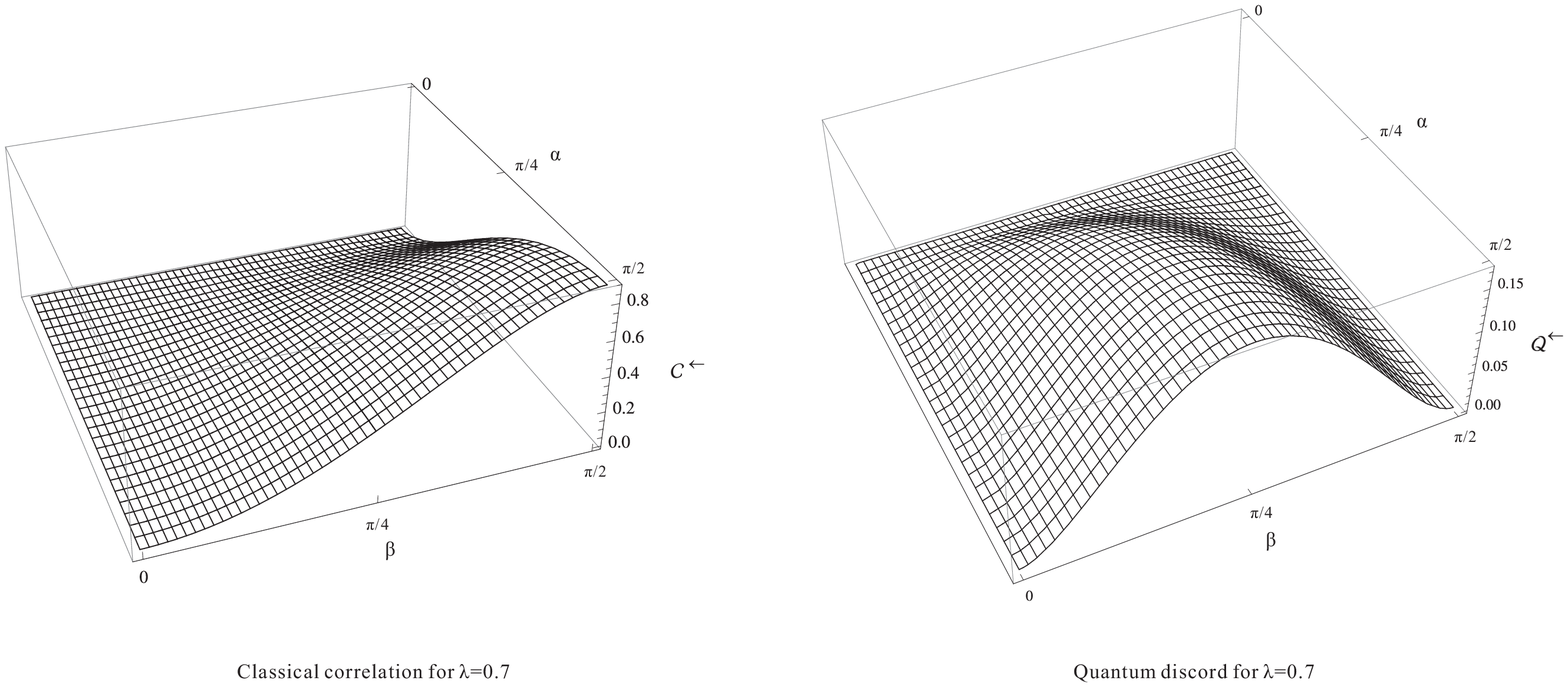}
\end{center}
 \caption{Classical correlation $\mathcal{C}^{\leftarrow}$ and quantum discord $\mathcal{Q}^{\leftarrow}$ of the states given by (\ref{mixture of two PP}) with $\lambda=0.7$.} \label{fig:CandQ2}
\end{figure}

\begin{figure}[tbph]
\begin{center}
  \includegraphics[width=0.8\textwidth]{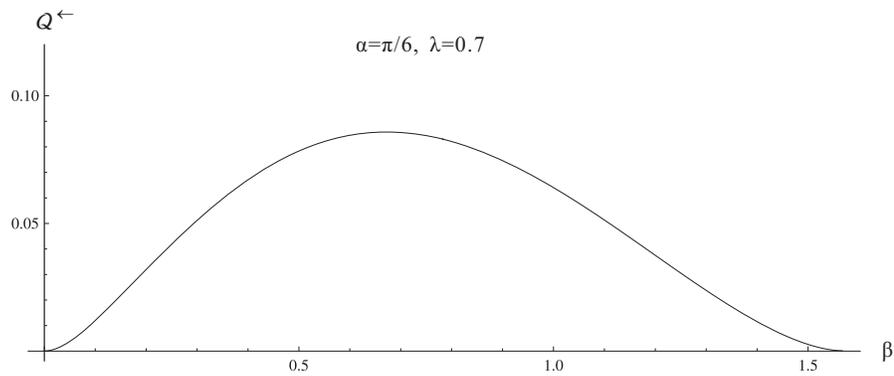}
\end{center}
\caption{Quantum discord $\mathcal{Q}^{\leftarrow}$ of states given by (\ref{mixture of two PP}) with $\alpha=\pi/6$ and $\lambda=0.7$. It is not symmetric with respect to $\beta\in[0,\frac{\pi}{2}].$} \label{fig:CandQ3}
\end{figure}

In figure \ref{fig:CandQ1} and figure \ref{fig:CandQ2}, we plot the classical correlation $\mathcal{C}^{\leftarrow}$ and quantum discord $\mathcal{Q}^{\leftarrow}$ for the states given by (\ref{mixture of two PP}) with $\lambda=0.5$ and $\lambda=0.7$ respectively. They are very similar to each other, but it should be noted that the plot of $\mathcal{Q}^{\leftarrow}$ in figure \ref{fig:CandQ1} is symmetric with respect to the parameter $\beta$, while it is not the case in \ref{fig:CandQ2}. To see this, refer to figure \ref{fig:CandQ3}.

It can be seen that among all states given by (\ref{mixture of two PP}) with fixed $\lambda$, the one with maximal classical correlation is of the form
\begin{equation*}
    \lambda\ket{0}\bra{0}\otimes\ket{0}\bra{0}
      +(1-\lambda)\ket{1}\bra{1}\otimes\ket{1}\bra{1},
\end{equation*}
and the one with maximal quantum discord is given by
\begin{equation*}
    \lambda\ket{0}\bra{0}\otimes\ket{0}\bra{0}
    +(1-\lambda)\ket{1}\bra{1}\otimes\ket{+}\bra{+}.
\end{equation*}
It is not difficult to see that the choice of $\lambda=1/2$ renders the largest classical correlation and quantum discord.

\section{More general states}

It is desirable to apply the geometric picture for a wider class of states.
This section is devoted to extending the discussion in Section 4 about X states to a more general case. Recall that in the geometric picture of X states Alice local state $\rho^A$ is located on the $y_3$ (or $y'_3$) axis, which is one of the symmetric axes of the ellipsoid $\mathfrak{E}$ (see figure \ref{fig:ellipsoid}). How about the case that point $A$ deviates from $y_3$ axis? We will discuss in this section this type of states. To begin with, let's see an example.

The example we will consider comes from \cite{Synak2004JPA}. Suppose a two-qubit pure state is given by \cite{note}
\begin{equation*}
    \ket{\psi^{AB}}=\frac{1}{\sqrt2}\big(\ket{\psi_0}_A\ket{0}_B
                    +\ket{\psi_1}_A\ket{1}_B),
\end{equation*}
where $\ket{\psi_0}_A=\frac{1}{\sqrt2}(\ket{0}+\ket{1})$ and $\ket{\psi_1}_A=\frac{4}{5}\ket{0}+\frac{3}{5}\ket{1}$.
Let qubit A pass through a quantum channel, the Kraus operators of which are given by
\begin{equation*}
    A_1=\ket{0}\bra{0}+\frac{1}{\sqrt2}\ket{1}\bra{1}, \quad
    A_2=\frac{1}{\sqrt2}\ket{0}\bra{1}.
\end{equation*}
Then output state of the channel is
\begin{equation*}
    \rho^{AB}
    =\sum_{i=1}^2(A_i\otimes\mathbbm1)\,\ket{\psi^{AB}}\bra{\psi^{AB}}\,
           (A_i\otimes\mathbbm1)^\dag.
\end{equation*}
Note that $\rho^{AB}$ here is not an X state.
With Bob performing measurement, the steering ellipsoid is given by
\begin{equation}\label{ellipsoid of Syntak}
    y_1^2+y_2^2+2\Big(y_3-\frac{1}{2}\Big)^2=\frac{1}{2}.
\end{equation}
The Bloch vector of $\rho^A$ is
\begin{equation*}
    \vec{r}^A=\bigg(\frac{49}{50\sqrt{2}},\;0,\;\frac{57}{100}\bigg).
\end{equation*}

\begin{figure}
\begin{center}
  \includegraphics[width=0.8\textwidth]{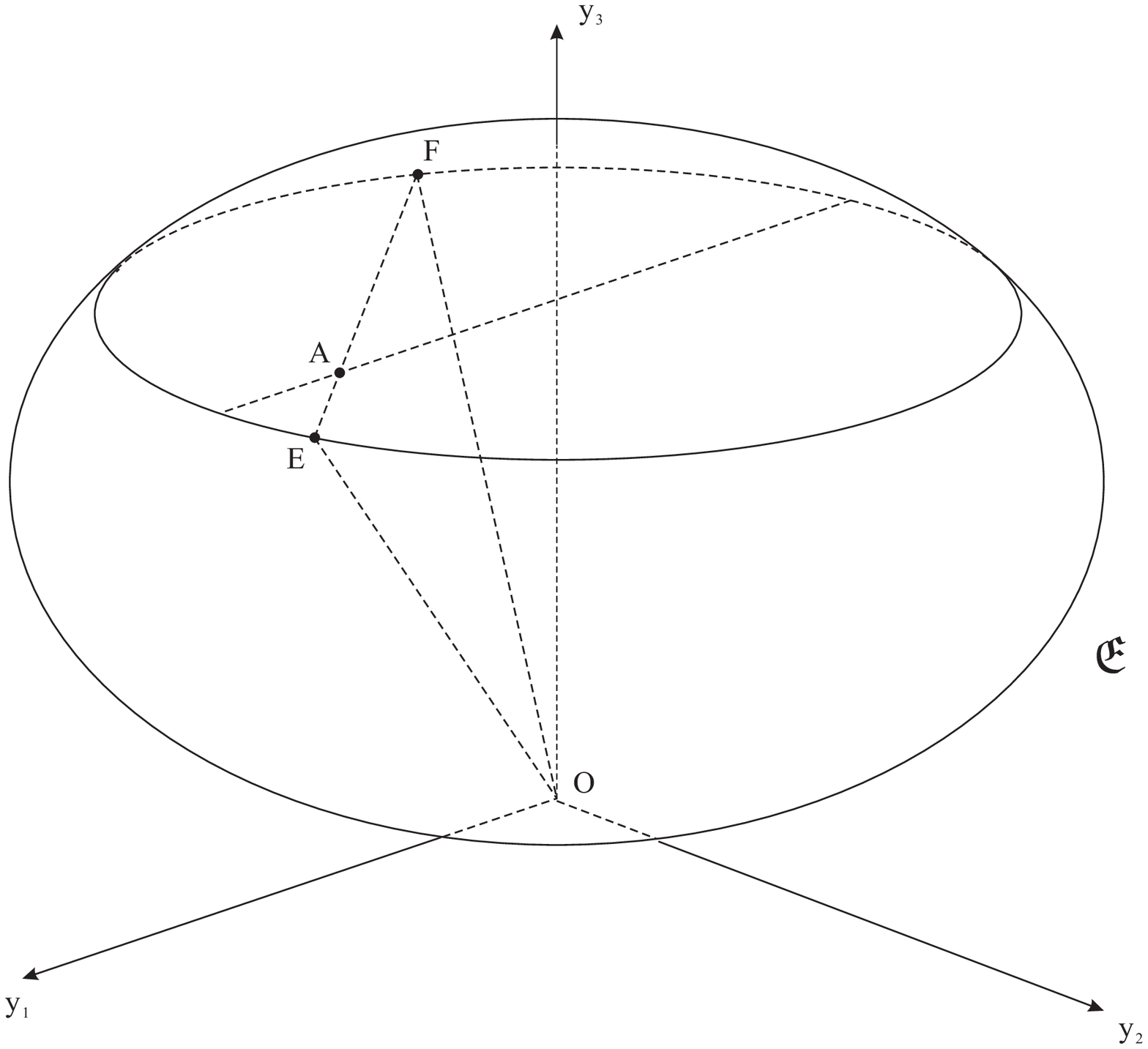}
\end{center}
\caption{The ellipsoid $\mathfrak{E}$ given by (\ref{ellipsoid of Syntak}) is cut by a plane that is parallel to $y_1\,y_2$ plane and contains point $A$. The intersection is a circle. Passing through point $A$, any chord of the circle leads to the optimal ensemble of $\rho^A$.}\label{fig:synak}
\end{figure}

The ellipsoid (\ref{ellipsoid of Syntak}) is symmetric under rotation about $y_3$ axis. The form is similar to that for X states. However, the vector $\vec{r}^A$ does not lie on $y_3$ axis (see figure \ref{fig:synak}). We can not obtain the $\overline{S}_{\min}^A$ analytically, but numerical evaluation reveals an interesting result: Considering any line passing through point $A$ and intersecting the ellipsoid at two point $E$ and $F$, the minimal value of $\overline{S}^A$ is reached at such $E$ and $F$ that $OE=OF$, or,
\begin{equation*}
    \overline{S}_{\min}^A=p_ES_E+p_FS_F=S_E=S_F,
\end{equation*}
where $S_E$ or $S_F$ have analytical expressions. We can see that it is the equi-entropy decomposition. Geometric description is clearly demonstrated in figure
\ref{fig:synak}.

The numerical value of $\overline{S}_{\min}^A$ is equal to $0.2804$, and classical correlation and quantum discord is given by
\begin{equation*}
    \mathcal{C}^{\leftarrow}=0.0118, \quad \mathcal{Q}^{\leftarrow}=0.0338.
\end{equation*}

The above example motivates us to consider more general two-qubit states. That is, the quantum steering ellipsoid can be translated along $y_3$ axis and no rotation is allowed. The position of local state $\rho^A$ is restricted in $y_1\,y_3$ plane.

We take into consider the states $\rho^{AB}$ with the following $R$ matrix.
\begin{equation*}
    R=\left(
    \begin{array}{cccc}
      1 & s_1 & 0 & s_3 \\
      r_1 & t_{11} & 0 & t_{13} \\
      0 & 0 & t_{22} & 0 \\
      r_3 & t_{31} & 0 & t_{33}
    \end{array}
      \right).
\end{equation*}
It is assumed that $R$ is non-singular, namely, $\det(R)\neq0$.
Obviously, the Bloch vector of $\rho^A$ is given by $\vec{r}^A=(r_1,\;0,\;r_3)$. Assuming that $s_1,t_{13}\neq 0$, we choice the parameters $t_{11}$ and $t_{33}$ as
\begin{equation*}
    t_{11}=\frac{r_1-s_3t_{13}}{s_1}, \quad
    t_{33}=\frac{r_1r_3s_1-r_1t_{31}+s_3t_{13}t_{31}}{s_1t_{13}}.
\end{equation*}
Under these conditions, we construct the states $\rho^{AB}$ randomly. The quantum steering ellipsoid $\mathfrak{E}$ given by
\begin{equation*}
    \frac{y_1^2}{\ell_1^2}+\frac{y_2^2}{\ell_2^2}
    +\frac{(y_3-Y_3)^2}{\ell_3^2}=1,
\end{equation*}
where
\begin{eqnarray*}
  & \ell_1^2=\frac{r_1^2 (1-s_1^2)-2 r_1s_3t_{13}+(s_1^2+s_3^2)t_{13}^2}
           {s_1^2(1-s_1^2-s_3^2)}, \\
  & \ell_2^2=\frac{t_{22}^2}{1-s_1^2-s_3^2}, \\
  & \ell_3^2=\frac{[r_1^2(1-s_1^2)-2r_1s_3t_{13}+t_{13}^2(s_1^2+s_3^2)]
                   (r_3s_1-t_{31})^2}{s_1^2t_{13}^2(1-s_1^2-s_3^2)^2}, \\
  & Y_3=\frac{r_3s_1t_{13}-r_1s_3(r_3s_1-t_{31})-t_{13}t_{31}(s_1^2+s_3^2)}
             {s_1t_{13}(1-s_1^2-s_3^2)}.
\end{eqnarray*}

Consider a class of lines passing through point $A$ and intersecting the $\mathfrak{E}$ at point $M$ and $N$. It follows that $\rho^A=p_M\rho_M+p_N\rho_N$. Corresponding this decomposition of $\rho^A$, the average entropy is given by $\overline{S}^A_{MN}=p_MS_M+p_NS_N$. We will find the minimal $\overline{S}^A$ among all these lines. Numerical results can be classified into the following two categories.

\textit{Class I} --- The optimal line is parallel to $y_1$ axis or $y_2$ axis. Denote by $E$ and $F$ the intersection points with $\mathfrak{E}$. The minimal $\overline{S}^A$ is given by $\overline{S}_{\min}^A=S_E=S_F$, meaning that it is an equi-entropy decomposition. Moreover, let $A'$ be the point that is on the line segment $EF$ and symmetric to point $A$ about $y_3$ axis. Then any point $A''$ between the point $A$ and $A'$ has the same minimal value of average entropy, namely,
\begin{equation*}
    \overline{S}^{A''}_{\min}=S_E=S_F\quad \mathrm{for}\quad  A''\in AA'.
\end{equation*}

\textit{Class I$\!$I} --- The optimal line is not parallel to $y_1\,y_2$ plane. In this case, let's consider the point $\tilde{A}$ with the coordinate $(0,\;0,\;r_3)$. Point $\tilde{A}$ in fact corresponds to the projection of the vector $\vec{r}^A$ onto $y_3$ axis.
We find that the minimal value of the average entropy $\overline{S}^{\tilde{A}}$ is given by the two points $G$ and $H$, which are the upper and lower apex of the ellipsoid $\mathfrak{E}$ respectively, that is,
\begin{equation*}
    \overline{S}_{\min}^{\tilde{A}}=p_GS_G+p_HS_H.
\end{equation*}
It is a quasi-eigendecomposition.

In the case of Class I, $\overline{S}^A_{\min}$ have analytical expression.
For Class I$\!$I, we only see that it has relationship with quasi-eigendecomposition. More effort is necessary to acquire further insight.

\section{Conclusion}

We present a geometric method as to how to describe and evaluate the minimal average entropy, which is the major obstacle in the computation of classical correlations and quantum discord.
For two-qubit states, the available ensemble of postmeasurement states of qubit A, which comes from the measurements performed on qubit B, is restricted in the quantum steering ellipsoid. The optimal ensemble can only be found on the surface of the ellipsoid.

For two-qubit X states, the geometric method provides a clear picture as well as exact results. We show that for X states the optimal decomposition is alternative: equi-entropy decomposition or quasi-eigendecomposition. In the geometric picture, equi-entropy decomposition corresponds to a horizontal line segment, while the quasi-eigendecomposition to a vertical one.
When an X state passing through some quantum channels, the dynamics of classical correlation and quantum discord can be easily analyzed in the geometric picture.

We extend the discussion about X states to a more general case by relaxing the requirement that the reduced density matrices are of diagonal form.
Little is known about the classical correlations or quantum discord of these states.
We perform numerical computations. A consequence of the numerical results is the following interesting alternative: the optimal decomposition is either equi-entropy decomposition, or is intimately related to the quasi-eigendecomposition. Combining these numerical results with the exact results for X states, we think that the geometric method may be generalized analytically rather than numerically. Further thought in this direction might be worthwhile.

The geometric viewpoint presented in this paper offers an alternative way to interpret and compute classical correlations and quantum discord. It is also useful in elucidating issues related to decoherence.
The remaining problem is to verify the conclusions we draw by means of numerical method. If these conclusions are indeed true, they will be helpful to work out the exact results of quantum discord analytically.

\bigskip

This work was supported by National Nature Science Foundation of China, the CAS, and the National
Fundamental Research Program 2007CB925200.

\clearpage
\end{document}